\documentstyle[aps,epsf,multicol]{revtex}
\begin{document}
\draft
\title{Universal Behavior of Load Distribution in Scale-free Networks}
\author{K.-I. Goh, B. Kahng, and D. Kim}
\address{School of Physics and Center for Theoretical Physics, 
Seoul National University, Seoul 151-747, Korea}
\date{\today}
\maketitle
\thispagestyle{empty}
\begin{abstract}
We study a problem of data packet transport in scale-free networks 
whose degree distribution follows a power-law with the exponent 
$\gamma$. Load, or betweenness centrality of a vertex is the 
accumulated total number of data packets passing through that 
vertex when every pair of vertices send and receive a data packet 
along the shortest path connecting the pair. It is found that 
the load distribution follows 
a power-law with the exponent $\delta \approx 2.2(1)$, 
insensitive to different values 
of $\gamma$ in the range, $2 < \gamma \le 3$, and different mean 
degrees, which is valid for both undirected and directed cases. 
Thus, we conjecture that the load exponent is a universal quantity 
to characterize scale-free networks. 
\end{abstract}
\pacs{PACS numbers: 89.70.+c, 89.75.-k, 05.10.-a}
\begin{multicols}{2}
\narrowtext
Complex systems consist of many constituents such as individuals, 
substrates, and companies in social, biological, and economic systems, 
respectively, showing cooperative phenomena between constituents 
through diverse interactions and adaptations to the pattern they 
create\cite{nature,science}. 
Interactions may be described in terms of graphs, consisting of 
vertices and edges, where vertices (edges) represent the constituents 
(their interactions). This approach was initiated by Erd\"os 
and R\'enyi (ER)\cite{er}. 
In the ER model, the number of vertices is fixed, while edges 
connecting one vertex to another occur randomly with certain 
probability. However, the ER model is too random to describe real 
complex systems. 
Recently, Watts and Strogatz (WS) \cite{ws} introduced a small-world 
network, where a fraction of edges on a regular lattice is rewired with 
probability $p_{\it WS}$ to other vertices. 
More recently, Barab\'asi and Albert (BA) \cite{ba,boston,porto} introduced 
an evolving network where the number of vertices $N$ increases linearly 
with time rather than fixed, and a newly introduced vertex is 
connected to $m$ already existing vertices, following the so-called 
preferential attachment (PA) rule.
When the number of edges $k$ incident upon a vertex is called the 
degree of the vertex, the PA rule means that the probability for 
the new vertex to connect to an already existing vertex is 
proportional to the degree $k$ of the selected vertex.
Then the degree distribution $P_D(k)$ follows a power-law 
$P_D(k)\sim k^{-\gamma}$ with $\gamma=3$ for the BA model, 
while for the ER and WS models, it follows a Poisson distribution. 
Networks whose degree distribution follows a power-law, called 
scale-free (SF) networks\cite{physica}, are ubiquitous in real-world networks 
such as the world-wide web\cite{www1,www2,www3}, the 
Internet\cite{internet1,internet2,internet3}, 
the citation network\cite{citation} 
and the author collaboration network of scientific papers 
\cite{co1,co2,co3}, and the metabolic networks in biological 
organisms\cite{metabolic}. 
On the other hand, there also exist random networks such as 
the actor network whose degree distribution follows a power-law 
but has a sharp cut-off in its tail\cite{class}. 
Thus, it has been proposed that the degree distribution can be 
used to classify a variety of diverse real-world networks\cite{class}. 
In SF networks, one may wonder if the exponent $\gamma$ 
is universal in analogy with the theory of critical phenomena; 
however, the exponent $\gamma$ turns out 
to be sensitive to the detail of network structure. 
Thus, a universal quantity for SF networks is yet to be found.
From a theoretical viewpoint, it is important to find a 
universal quantity for SF networks, which is a purpose 
of this Letter.

A common feature between the WS and SF networks would be 
the small-world property that the mean separation between two 
vertices, averaged over all pairs of vertices (called the diameter hereafter), 
is shorter than that of a regular lattice. The small-world property in 
SF networks results from the presence of a few vertices with high degree. 
In particular, the hub, the vertex whose degree is the largest, plays a 
dominant role in reducing the diameter of the system. 
Diameters of many complex networks in real world are small, 
allowing objects transmitted through the network such as neural 
spikes on neural network, or data packets on Internet,  
to travel from one vertex to another quickly along the shortest path.
The shortest paths are indeed of relevance to network transport properties. 
When a data packet is sent from one vertex to another through SF 
networks such as Internet, it is efficient to take a road along 
the shortest paths between the two. 
Then vertices with higher degrees should be heavily loaded and jammed 
by lots of data packets passing along the shortest paths.  
To prevent such Internet traffic congestions, and allow data 
packets to travel in a free-flow state, one has to enhance the 
capacity, the rate of data transmission, of each vertex to the 
extent that the capacity of each vertex is large enough to handle 
appropriately defined ``load''. 

In this Letter, we define and study such a quantity, which we simply 
call load, to characterize the transport dynamics in SF networks. 
In fact, this quantity turns out to be equivalent to ``betweenness 
centrality" which was introduced in a social network to quantify 
how much power is centralized to people in social networks
\cite{co2,social}. 
While it has been noted that the betweenness centrality has 
a long tail\cite{private}, here we focus our attention 
on its probability distribution for various SF networks with 
different degree exponents. 
Thus knowing the distribution of such quantity enables us to not only 
estimate the capacity of each vertex needed for a free-free state, 
but also understand the power distribution in social networks, 
which is another purpose of this Letter. 

To be specific, we suppose that a data packet is sent from a 
vertex $i$ to $j$, for every ordered
pair of vertices $(i,j)$. For a given pair $(i,j)$, it 
is transmitted along the shortest path between them. 
If there exist more than one shortest paths, the data packet 
would encounter one or more branching points. 
In this case, we assume that the data packet is divided evenly by 
the number of branches at each branching point as it travels.
Then we define the load $\ell_k$ at a vertex $k$ as the 
total amount of data packets passing through that vertex $k$ 
when all pairs of vertices send and receive one unit of data packet
between them. Here, we do not take into account the time delay 
of data transfer at each vertex or edge, so that all data are 
delivered in a unit time, regardless of the distance between 
any two vertices.  

We find numerically that the load 
distribution $P_L(\ell)$ follows a power-law $P_L(\ell)\sim \ell^{-\delta}$.
Moreover, the exponent $\delta\approx 2.2$ we obtained is 
insensitive to the detail of the SF network structure as long 
as the degree exponent is in the range, $2 < \gamma \le 3$.  
The SF networks we used do not permit rewiring process, and 
the number of vertices are linearly proportional to that of 
edges. When $\gamma > 3$, $\delta$ increases as $\gamma$ 
increases, however. The universal behavior is valid for 
directed networks as well, 
when $2 < \{\gamma_{\rm in}, \gamma_{\rm out} \} \le 3$.
Since the degree exponents in most of real-world SF networks 
satisfy $2 < \gamma \le 3$, the universal behavior is interesting. 
   
We construct a couple of classes of undirected SF networks both 
in the static and evolving ways. Each class of networks include 
a control parameter, 
according to which the degree exponent is determined. 
First, we deal with the static case. There are $N$ vertices in the 
system from the beginning, which 
are indexed by an integer $i$ $(i=1,\dots,N)$. We assign the weight 
$p_i = i^{-\alpha}$ to each vertex, where $\alpha$ is a control parameter
in $[0,1)$. Next, we select two different vertices $(i,j)$ 
with probabilities equal to the normalized weights, $p_i/\sum_k p_k$ 
and $p_j/\sum_k p_k$, respectively, and add an edge between them 
unless one exists already. 
This process is repeated until $mN$ edges are made in the system. 
Then the mean degree is $2m$.
Since edges are connected to a vertex with frequency proportional 
to the weight of that vertex, the degree at that vertex is given as  
\begin{equation}
{{k_i} \over {\sum_j k_j}}\approx 
\frac{(1-\alpha)}{N^{1-\alpha}i^{\alpha}},  
\end{equation}
where $\sum_j k_j=2mN$. Then it follows that the degree distribution 
follows the power-law, $P_D(k)\sim k^{-\gamma}$, where $\gamma$ 
is given by 
\begin{equation} 
\gamma = (1+\alpha)/\alpha. 
\end{equation}
Thus, adjusting the parameter $\alpha$ in [0,1), we can obtain various 
values of the exponent $\gamma$ in the range, $2 < \gamma < \infty$. 

Once a SF network is constructed, we select an ordered pair of 
vertices $(i,j)$ on the network, 
and identify the shortest path(s) between them and measure 
the load on each vertex along the shortest path using the 
modified version of the breath-first search algorithm 
introduced by Newman \cite{co2}. 
It is found numerically that the load $\ell_i$ at vertex $i$ 
follows the formula,  
\begin{equation}
{{\ell_i} \over {\sum_j \ell_j}}\sim 
\frac{1}{N^{1-\beta}i^{\beta}},  
\end{equation}
with $\beta=0.80(5)$. This value of $\beta$ is
insensitive to different values of the exponent $\gamma$ 
in the range, $2 < \gamma \le 3$ as shown in the inset of Fig.~1. 
The total load, $\sum_j \ell_j$ scales 
as $\sim N^2 \log N$. This is because there are $N^2$ pairs 
of vertices in the system and the sum of the load contributed 
by each pair of vertices is equal to the distance between the two 
vertices, which is proportional to $\log N$. 
Therefore, the load $\ell_i$ at a vertex 
$i$ is given as 
\begin{equation}
\ell_i \sim (N \log N) ({N}/{i})^{\beta}.
\end{equation}
From Eq.(4), it follows that the load distribution scales as 
$P_L(\ell) \sim \ell^{-\delta}$, with $\delta = 1+1/\beta \approx 2.2(1)$,
independent of $\gamma$ in the range, $2 < \gamma \le  3$. 
Direct measure of $P_L(\ell)$ also gives $\delta \approx 2.2(1)$ 
as shown in Fig.~1. We also check $\delta$ for different mean degrees 
$m=2,4$ and 6, but obtain the same value, $\delta \approx 2.2(1)$. 
Thus, we conclude that the exponent $\delta$ is a generic quantity 
for this network. Note that Eqs.(1) and (4) combined gives a scaling 
relation between the load and the degree for this network as 
\begin{equation}
\ell \sim k^{(\gamma-1)/(\delta-1)}.
\end{equation}
Thus, when and only when $\gamma=\delta$, the load at each vertex
is directly proportional to its degree. Otherwise, it scales 
nonlinearly. On the other hand, for $\gamma >3$, the exponent 
$\delta$ depends on the exponent $\gamma$ in a way that it increases 
as $\gamma$ increases. Eventually, the load distribution decays 
exponentially for $\gamma = \infty$ as shown in Fig.~1. 
Thus, the transport properties of the SF networks with $\gamma >3$ are 
fundamentally different from those with $2 < \gamma \le 3$. This is 
probably due to the fact that for $\gamma >3$, the second moment 
of $P_D(\ell)$ exists, while for $ \gamma \le 3$, it does not.

We examine the system-size dependent behavior of the load at 
the hub, $\ell_h$ for the static model. According to Eq.(4), 
$\ell_h$ behaves as $\ell_h \sim N^{1.8}\log N$ in the range, 
$2 < \gamma \le 3$, while for $\gamma >3$, $\ell_h$ increases 
with $N$ but at a much slower rate than that for 
$2 < \gamma \le 3$ as shown in Fig.~2. 
That implies that the shortest pathways between two vertices 
become diversified, and they do not necessarily pass through 
the hub for $\gamma >3$. That may be related to the result 
that epidemic threshold is null in the range $2 < \gamma \le 3$, 
while it is finite for $\gamma > 3$ in SF networks, because 
there exist many other shortest paths not passing through 
the hub for $\gamma >3$, so that the infection of the hub does 
not always lead to the infection of the entire system. 
Thus, epidemic threshold is finite for $\gamma >3$ \cite{epidemic}.    


Next, we generate other SF networks in an evolving way, 
using the methods proposed by Kumar {\it et al.}\cite{copying}, 
and by Dorogovtsev {\it et al.}\cite{porto}. In these cases, we 
also find the same results as in the case of static models. 

We also consider the case of directed SF network. The directed SF 
networks are generated following the static rule. 
In this case, we assign two weights $p_i=i^{-\alpha_{\rm out}}$ and 
$q_i=i^{-\alpha_{\rm in}}$ $(i=1,\dots,N)$ to each vertex for outgoing 
and incoming edges, respectively. Both control parameters 
$\alpha_{\rm out}$ and $\alpha_{\rm in}$ are in the interval $[0,1)$. Then two 
different vertices $(i,j)$ are selected with probabilities, 
$p_i/\sum_k p_k$ and $q_j/\sum_k q_k$, respectively, and an edge 
from the vertex $i$ to $j$ is created with an arrow, $i \rightarrow j$. 
The SF networks generated in this way show the power-law in 
both outgoing and incoming degree distributions with the 
exponents $\gamma_{\rm out}$ and $\gamma_{\rm in}$, respectively. 
They are given as $\gamma_{\rm out}=(1+\alpha_{\rm out})/\alpha_{\rm out}$ and 
$\gamma_{\rm in}=(1+\alpha_{\rm in})/\alpha_{\rm in}$. 
Thus, choosing various values of $\alpha_{\rm out}$ and $\alpha_{\rm in}$, we 
can determine different exponents $\gamma_{\rm out}$ and $\gamma_{\rm in}$.  
Following the same steps as for the undirected case, we obtain 
the load distribution on the directed SF networks. The load 
exponent $\delta$ obtained is $\approx 2.3(1)$ as shown in Fig.~3, 
consistent with the one for the undirected case, also being 
independent of $\gamma_{\rm out}$ and $\gamma_{\rm in}$ in 
$2 < \{ \gamma_{\rm out}, \gamma_{\rm in} \} \le 3$. Therefore, 
we conjecture that the load exponent is a universal value 
for both the undirected and directed cases.

To see if such universal value of $\delta$ appears in real world 
network, we analyzed the co-authorship network, where nodes represent 
scientists and they are connected if they wrote a paper together. 
The data are collected in the field of the neuro-science, published 
in the period 1991-1998\cite{co3}. This network is appropriate 
to test the load, $i.e.$, the betweenness centrality distribution, 
because it does not include rewiring process as it evolves, and 
its degree exponent $\gamma \approx 2.2$ lies 
in the range $2 < \gamma \le 3$. As shown in Fig.~4, the load 
distribution follows a power-law with the exponent $\delta \approx 
2.2$, in good agreement with the value obtained in the previous 
models. 

We also check the load distribution for the case when data 
travel with constant speed, so that the time delay of data 
transfer is proportional to the distance between two vertices. 
We find that the time delay effect does not change the load 
distribution and the conclusion of this work. The reason of this 
result is that when the time delay is accounted, load at each vertex 
is reduced roughly by a factor $\log N$, proportional to 
the diameter, which is negligible compared with the load without 
the time delay estimated to be $\sim N^{1.8}\log N$ in Eq.(4). 
Due to this small-world property, the universal behavior remains 
unchanged under the time delay of data transmission. 

Finally, we mention the load distribution of the small-world 
network of WS which is not scale-free. It is found that its load 
distribution does not obey a power-law, but shows a combined behavior 
of two Poisson-type decays resulting from short-ranged 
and long-ranged connections, respectively, as shown in Fig.~5. 
We also find the average load, 
${\bar \ell}(p_{\it WS})\equiv (1/N)\sum_i \ell_i(p_{\it WS})$, as a 
function of the rewiring probability $p_{\it WS}$ decays rapidly 
with increasing $p_{\it WS}$, behaving similar to the diameter in 
the WS model, as shown in the inset of Fig.~5.

In conclusion, we have considered a problem of data packet transport 
on scale-free networks generated according to preferential attachment 
rules and introduced a physical quantity, 
load $\{\ell_i \}$ at each vertex. 
We found that the load distribution 
follows a power-law, $P_L(\ell)\sim \ell^{-\delta}$, with 
the exponent $\delta \approx 2.2(1)$, which turns out to be 
insensitive to the degree exponent in the range $(2, 3]$ 
when rewiring process is not included and the networks are of  
unaccerelated growth. 
Moreover, it is also the same for both directed and undirected cases
within our numerical uncertainties. 
Therefore, we conjecture that the load exponent is a generic quantity 
to characterize scale-free networks. 
The universal behavior we found may have interesting implications 
to the interplay of SF network structure and dynamics. 
For $\gamma >3$, however, the 
load exponent $\delta$ increases as the degree exponent 
$\gamma$ increases, and eventually the load distribution decays 
exponentially as $\gamma \rightarrow \infty$. 
It would be interesting to examine the robustness of the universal 
behavior of the load distribution under some modifications of 
generating rules for SF networks such as rewiring process and 
acceleration growth, which, however, is beyond the scope of the 
current study.   

We would like to thank M.E.J. Newman for introducing Refs.
\cite{co2,social,private}, 
and H. Jeong for providing the data for the co-authorship network.
This work is supported by grants No.2000-2-11200-002-3 from the 
BRP program of the KOSEF. 
%
%

%
%
\begin{figure}
\centerline{\epsfxsize=8cm \epsfbox{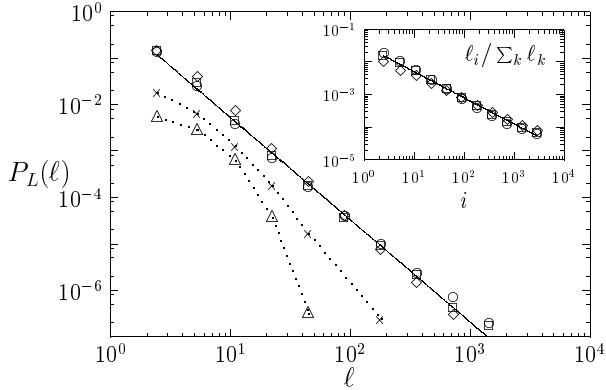}}
\caption{Plot of the load distribution $P_L(\ell)$ versus $\ell$ for 
various $\gamma=2.2$ ($\circ$), 2.5 ($\Box$), 3.0 ($\diamond$), 
4.0 ($\times$), and $\infty$ ($\triangle$) in double logarithmic scales. 
The linear fit (solid line) has a slope $-2.2$. 
Data for $\gamma > 3.0$ are shifted vertically for clearance.
Dotted lines are guides to the eye.
Simulations are performed for $N=10,000$ and $m=2$ and all data 
points are log-binned, averaged over 10 configurations. Inset: 
Plot of the normalized load $\ell_i/\sum_k \ell_k$ 
versus vertex index $i$ in double logarithmic scales 
for various $\gamma=2.2$ ($\circ$), 2.5 ($\Box$),  
and 3.0 ($\diamond$).} 
\label{fig1}
\end{figure}

\begin{figure}
\centerline{\epsfxsize=8cm \epsfbox{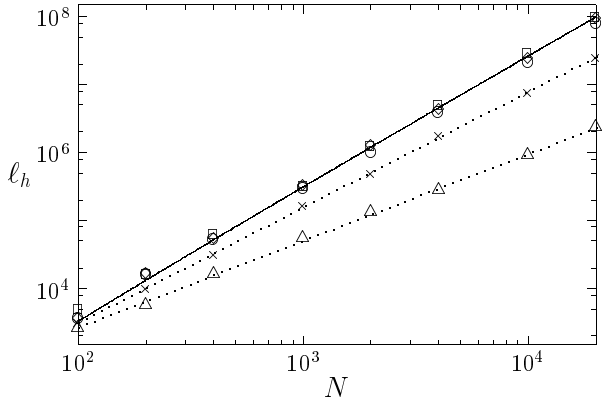}}
\caption{Plot of the system-size dependence of the load at the 
hub versus system size $N$ for various $\gamma=2.2$ ($\circ$),
2.5 ($\Box$), 3.0 ($\Diamond$), 4.0 ($\times$), 
and $\infty$ ($\triangle$).
Solid line is $N^{1.8}\log N$ and dotted lines have slopes 
1.70 and 1.25, respectively, from top to bottom.
Simulations are performed for $m=2$ and 
all data points are averaged over 10 configurations.}
\label{fig2}
\end{figure}
\begin{figure}
\centerline{\epsfxsize=8cm \epsfbox{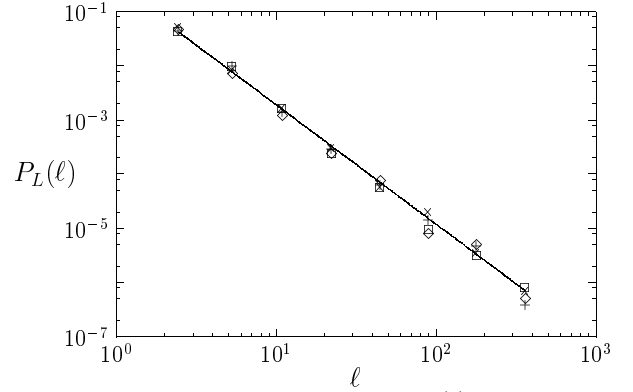}}
\caption{Plot of the load distribution $P_L(\ell)$ versus 
$\ell$ for the directed case. The data are obtained for 
$(\gamma_{\rm in}, \gamma_{\rm out}) =$ (2.1, 2.3) ($\diamond$), 
(2.1, 2.7) ($+$), (2.5, 2.7) ($\Box$) and (2.5, 2.2) ($\times$). 
The fitted line has a slope $-2.3$. All data points are log-binned.} 
\label{fig3}
\end{figure}

\begin{figure}
\centerline{\epsfxsize=7cm \epsfbox{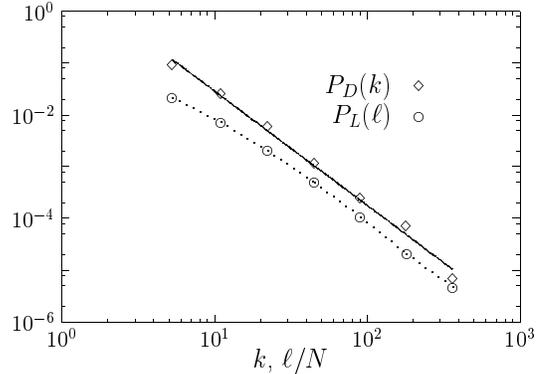}}
\caption{Plot of the degree distribution $P_D(k)$ ($\Diamond$) and the 
load distribution $P_L(\ell)$ ($\circ$) for a real-world network, 
the co-authorship network. The number of vertices 
(different authors) are 205,202. Least square fit (solid line) has
a slope $-2.2$. All data points are log-binned.}
\label{fig4}
\end{figure}

\begin{figure}
\centerline{\epsfxsize=8cm \epsfbox{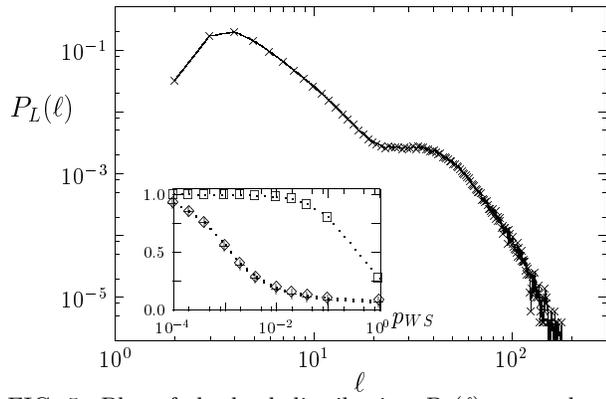}}
\caption{Plot of the load distribution $P_L(\ell)$ versus load 
$\ell$ for the small-world network. Simulations are performed 
for system size $N=1,000$, and average degree $\langle k \rangle =10$, 
and the rewiring probability $p_{\it WS}=0.01$, averaged over 500 
configurations. Inset: Plot of the average load ($\diamond$), 
diameter ($+$), clustering coefficient ($\Box$) versus the 
rewiring probability $p_{\it WS}$. All the data are normalized 
by the corresponding values at $p_{\it WS}=0$. Dotted lines are 
guides to the eye.}
\label{fig5}
\end{figure}

\end{multicols}
\end{document}